# Electroweak Model of Lepton Mass and Mixing Hierarchies

E. M. Lipmanov

40 Wallingford Road # 272, Brighton MA 02135, USA

**Abstract**

Flavor physics, like cosmology, is likely in need of new basic ideas; the puzzles of elementary particle mass hierarchies and in particular the e-mu-tau and neutrino ones still remain mysteries. In this paper a new idea of dynamical connection between low energy 3-flavor particle mass hierarchies and electroweak charges is studied with restriction to the simplest case of lepton flavor phenomenology. The main inference is that it can be only two types of lepton 3-flavor particle-copy groups: 1) with large and strongly hierarchical mass ratios and 2) with close to 1 mass ratios. From experimental data definitely follows that the three charged leptons belong to the first type whereas the three neutrinos belong to the second type and so are quasi-degenerate. The inferences of QD-neutrinos with realistic small masses and oscillation hierarchy parameter and quark-QD-neutrino mixing angle complementarity follow from the fact of small EW charges and their relation to the concept of benchmark flavor pattern.

## 1. Introduction and motivation for the main idea

The factual contents of known flavor physics are particle mass matrices. Since the origin of the basic particle mass scale may be considered beyond flavor scope[1], the observable quantities of flavor physics are dimensionless ones such as mass ratios, mixing angles and CP-violating phases.

---

[1] Flavor physics is about particle mass copies and their relations; not about the origin of mass. The latter is commonly related to vacuum expectation values of scalar fields.

With restriction to lepton EW physics[2], there are only two other basic dimensionless quantities – dimensionless-made electric e and semi-weak $g_W$ charges. In known EW gauge theory they are not related to flavor particle mass quantities. The main motivations of present study are:

1) A completely new (unification) idea is considered -- EW charges are related to the substance of flavor physics in the sense that without extra particle generations there would be no EW interactions of the first generation particles. The contents of this suggestion are *dynamical[3] connections between EW charges and deviations from mass-degeneracy (DMD) of flavor particle-copies*. It is a phenomenological presentation of new physics that seems lurking behind the fact of extra particle generation in the electroweak theory (EWT) with EW charges as free parameters. That physical idea is suggestive and leads to testable quantitative inferences.

2) The highly successful one-generation lepton EWT [1] establishes two small EW charges of a SU(2)-doublet of electron and neutrino with masses $m_e$ and $m_v$. The EW fields are generated by local gauge symmetry via EW charges, but the magnitudes of these charges are free parameters subjected only to one condition $\neq 0$ (EW interactions must exist). The physical meaning of the EW charge magnitudes is definitely beyond the scope of the EWT. Another basic unexplained fact in the known realm of EW interactions is the physical meaning of elementary particle flavor degree of freedom (why cannot nature exist without it?). To connect these two principal problems is a leading motivation for the stated above idea.

3) Empirical fact of two extra particle copies (generations) that enter the Standard Model by free parameters is a known for a long time qualitative evidence of necessary new fundamental physics beyond the

---

[2] Restriction to the simplest case of lepton physics is encouraged by the great examples from physics history in particular at the formation of quantum mechanics.

[3] In contrast to known unifications in particle physics based on symmetry, the discussed new kind of unifications in EWT is called 'dynamical'.

SM – it is since symmetry cannot answer the physically meaningful and needed 'why' question. One-generation EWT semi-empirically extended to three generations [2], [3], [4], though physically motivated but without an answer of why not more, gains more than threefold enlarged number of free parameters. Substantially motivated reduction of the number of free parameters in EWT without weakening its predictive power is a topical problem. It is the third strong point of the suggested connection between EW-charges and flavor quantities.

4) Finally, the old plus recent empirical facts of extreme, large and small, lepton and quark mass and mixing hierarchies are in pressing need for phenomenological explanation (e.g. connection to generic EW quantities). It is a certain motivation for the above idea.

In Sec.2 the main idea is quantitatively stated. In the main Sec.3 a complete set of equations for lepton DMD-quantities with EW charges as sources is formulated, solved and inferences discussed. Sec.4 contains conclusions.

## 2. Electroweak theory and flavor degree of freedom

An apparent radical way to substantially reduce the number of dimensionless flavor parameters in the EWT is to connect them with the EW charges. This can be envisaged in low energy phenomenology only due to the empirical fact of flavor particle degree of freedom. With three flavor generations[4] the connections of EW charges with flavor quantities should be unique and so the comparison with experimental data should be decisive.

There are two DMD-quantities and one DMDH-hierarchy for three charged leptons (CL) and the same for three neutrinos ($\nu$):

$$\text{DMD(CL)}1 \equiv [(m_\tau^2/m_\mu^2)-1], \quad \text{DMD(CL)}2 \equiv [(m_\mu^2/m_e^2)-1], \quad (1)$$

$$\text{DMDH(CL)} \equiv \text{DMD(CL)}1 \,/\, \text{DMD(CL)}2; \quad (2)$$

$$\text{DMD}(\nu)1 \equiv [(m_2^2/m_1^2)-1], \quad \text{DMD}(\nu)2 \equiv [(m_3^2/m_2^2)-1], \quad (3)$$

$$\text{DMDH}(\nu) \equiv \text{DMD}(\nu)1 \,/\, \text{DMD}(\nu)2. \quad (4)$$

---

[4] The empirical low energy elementary particle island contains only three generations.

$m_e$, $m_\mu$ and $m_\tau$ are the CL masses and $m_1$, $m_2$, $m_3$ are organized three neutrino masses $m_1 < m_2 < m_3$. For simplicity, only 'normal' neutrino mass ordering is considered.

The guiding approach in this paper is dual to that of symmetry; the main quantities in the present model are *deviations from symmetry* that are related to mass degeneracy of elementary particles. So, in accordance with Sec.1, the quantitative contents of suggested idea are described by two analogous equations:

$$\text{DMDH}(\text{CL}) = \alpha(q^2 \cong M_Z^2), \quad \text{DMDH}(\nu) = \alpha_W(q^2 \cong M_Z^2), \qquad (5)$$

$\alpha$ is the fine structure constant and $\alpha_W = g_W^2/4\pi$ is the semi-weak analogue of the fine structure constant $\alpha$. These equations are considered at electroweak mass scale.

By relations (5), EW charges though not related to individual lepton masses are related to particle-copy *mass distributions* and thus to the system of three lepton generations in a substantial way that enhanced the unity of EWT.

### 3. Lepton flavor phenomenology from two basic physical premises

**1.** <u>Two physical premises</u> —

**a**) Dynamical relations (5) between CL and neutrino DMDH-hierarchies and electric and semi-weak charges,

**b**) Universal quadratic relations between generic flavor DMD-quantities (1 and 2), see [6]-[7], —
determine a complete universal system of two equations for four lepton mass-ratio DMD-quantities and lepton and quark mixing angle DMD-quantities with *EW charges as sources*. Consider separately these equations for lepton mass ratios and the mixing angles.

**2.** <u>Lepton mass ratios</u>. The two stated physical premises lead to two flavor equations:

$$[\text{DMD}(\text{CL or }\nu)1]/[\text{DMD}(\text{CL or }\nu)2] = (\alpha \text{ or } \alpha_W), \qquad (6)$$

$$[\text{DMD}(1 \text{ or } 2)]^2 = 2[\text{DMD}(2 \text{ or } 1)]. \qquad (7)$$

Notations for DMD-quantities and hierarchies are as in (1)-(4).

Since $\alpha$ and $\alpha_W$ are small in comparison with 1, from equation (6) follow conditions

$$\text{DMD}\,1 \ll \text{DMD}\,2. \qquad (8)$$

Namely in all cases the two DMD-quantities cannot be simultaneously of order 1; the *DMD-quantities in the pairs are strongly hierarchical* in all cases including CL and neutrino masses and mixing. With that remark and inferences from the quadratic hierarchy relations (7), there are only two options,

$$[\text{DMD}(1;2)] \gg 1 \text{ or } [\text{DMD}(1;2)] \ll 1. \qquad (9)$$

The results (8) and (9) have a clear physical meaning: the two flavor equations (6) and (7) predict that particle DMD-quantities are either very large or very small (against 1) with the DMDH-hierarchies (2) and (4) being always very large. The conditions (9) determine the order choice in the quadratic relations (7). That choice is substantial and should be resolved in each case from comparison with the experimental data.

From CL experimental data it obviously follows that the source in Eq.(6) is $\alpha$, not $\alpha_W$, with condition

$$\text{DMD}(\text{CL}) \gg 1. \qquad (10)$$

Thus, the pair of equations for CL mass ratios is uniquely given by

$$\text{DMDH}(\text{CL}) = \alpha(q^2 = M_Z^2), \quad [\text{DMD}(\text{CL})1]^2 = 2[\text{DMD}(\text{CL})2]. \qquad (11)$$

Solution of the pair of equations (11) for CL DMD-quantities and mass ratios is given by (at $q^2 = M_Z^2$)

$$[\text{DMD}(\text{CL})2] = 2/\alpha^2, \quad m_\mu/m_e = \sqrt{2}/\alpha, \qquad (12)$$

$$[\text{DMD}(\text{CL})1] = 2/\alpha, \quad m_\tau/m_\mu = \sqrt{(2/\alpha)}. \qquad (13)$$

From quantitative estimations with $\alpha(q^2 = M_Z^2) \cong 1/129$ the $m_\mu/m_e$ mass ratio (12) differs from data value [4] by ~10%, while $m_\tau/m_\mu$ mass ratio (13) - by ~4%.

So, the pair of equations (11) in lepton phenomenology answers the longstanding problems of 'why both muon and tau with masses in low energy region are needed?' – The answer from solutions (12)-(13) is '*if* $\alpha \to 0$, $m_\tau$ and $m_\mu$ would get unobservable heavy;

and so the muon and tau with realistic masses are needed for the existence of realistic EW interactions with meaningful connections between particle masses and charges[5'].

Analogous to the CL conditions (10) neutrino DMD-conditions $DMD(\nu) \gg 1$ are forbidden by the oscillation data – the choice $DMD(\nu) \gg 1$ would lead to not-QD neutrino type, but it is not acceptable since contradicts experimental data [8] on solar and atmospheric neutrino oscillation mass-squared differences and especially on oscillation hierarchy parameter – $r = \Delta m^2_{sol}/\Delta m^2_{atm} \cong \alpha_W^2/2 \cong 1/1800$ is in clear disagreement with data (the 3σ data ranges [8] are $r \cong 0.027 - 0.040$)[6].

And so, in the neutrino case the other condition must be true,

$$DMD(\nu) \ll 1, \quad m_3/m_2 \cong 1, \quad m_2/m_1 \cong 1. \tag{14}$$

This is the condition for quasi-degenerate neutrinos. The pair of equations for neutrino mass ratios is unambiguously given by

$$DMDH(\nu) = \alpha_W(q^2 = M_Z^2), \quad [DMD(\nu)2]^2 = 2[DMD(\nu)1]. \tag{15}$$

The important difference between CL and neutrino options[7] is reflected in the different ordering of [DMD1] and [DMD2] terms in the quadratic relations of (11) and (15). It leads to essential physical result that the neutrino DMD-quantities must be small and means that relations (14) and (15) *unambiguously predict QD-neutrino type from the condition of small EW charges and experimental data especially on the oscillation hierarchy parameter r*.

---

[5] In one-generation EWT with no mass copies the particle masses and charges cannot be related in any physical aspect – by definition; it is apparent even in EWT extended by extra generations [2]-[4], but it is not so in the studied here model.

[6] The fourth option (from equations (6) and (7)) is inverse ordering of the DMD-quantities (1) in the quadratic relation from (11). It leads to QD-mass spectrum with solution that follows from (16)-(17) after replacement $\alpha_W \to \alpha$. It cannot be related to neutrinos since it is in definite disagreement with oscillation data.

[7] In ref. [6] the striking difference between the magnitudes of DMD-quantities of CL and neutrinos is related to possible Majorana neutrino nature and defined there general Dirac-Majorana DMD-duality condition.

The solutions of the pair of equations (15) for QD-neutrino DMD-quantities are

$$[DMD(\nu)2] = [(m_3^2/m_2^2)-1] = \Delta m^2_{atm}/m_2^2 = 2\alpha_W(q^2=M_Z^2), \quad (16)$$

$$[DMD(\nu)1] = [(m_2^2/m_1^2)-1] = \Delta m^2_{sol}/m_1^2 = 2[\alpha_W(q^2=M_Z^2)]^2. \quad (17)$$

As conclusion, the absolute neutrino squared masses must be considerable larger than the neutrino oscillation mass-squared differences,

$$m_2^2 \cong \Delta m^2_{atm}/2\alpha_W, \quad m_1^2 \cong \Delta m^2_{sol}/2\alpha_W^2. \quad (18)$$

Quantitative estimations of QD-neutrino mass ratios

$$m_3/m_2 \cong \exp(-\alpha_W), \quad m_2/m_1 \cong \exp(-\alpha_W^2), \quad (19)$$

the magnitude of neutrino oscillation parameter

$$r = \Delta m^2_{sol}/\Delta m^2_{atm} \cong \alpha_W(q^2=M_Z^2) \cong 1/30 \quad (20)$$

and the absolute QD-neutrino masses (with best fit oscillation data [8])

$$m_2 \cong [\Delta m^2_{atm}/2\alpha_W]^{1/2} \cong [\Delta m^2_{sol}/2]^{1/2}/\alpha_W \cong m_1 \cong 0.19 \text{ eV} \quad (21)$$

are inferences from solutions (16) and (17); the phenomenon of neutrino oscillations is a necessary result of the neutrino weak interactions – if $\alpha_W = 0$, neutrinos would be exactly mass-degenerate with no oscillations.

**3. Neutrino mixing.** As next step, DMD-quantities for the solar $\theta_{12}$ and atmospheric $\theta_{23}$ neutrino mixing angles and the corresponding DMD-hierarchy must be considered. Since the dynamical sources of lepton flavor quantities are small EW charges, $\alpha, \alpha_W \ll 1$, the known from data large physical effect of neutrino mixing cannot be of dynamical origin. And so, one should introduce the concept of 'benchmark flavor pattern' [5] that among others contains the neutrino ($\nu$) and quark (q) benchmark mixing matrices[8],

$$\begin{pmatrix} 1 & 0 & 0 \\ 0 & 1 & 0 \\ 0 & 0 & 1 \end{pmatrix}_q, \quad \begin{pmatrix} 1/\sqrt{2} & 1/\sqrt{2} & 0 \\ -1/2 & 1/2 & 1/\sqrt{2} \\ 1/2 & -1/2 & 1/\sqrt{2} \end{pmatrix}_\nu. \quad (22)$$

---

[8] Without concepts of benchmark mixing level the DMD-quantities for neutrinos (24) and quark (26) cannot be properly defined.

The benchmark mixing is thought not dynamical[9], i.e. not determined by the EW charges; the dynamical mixing is determined by the EW charges and is always small.

Realistic DMD-quantities for the solar $\theta_{12}$ and atmospheric $\theta_{23}$ neutrino mixing angles and also quark mixing angles ($\theta_c$ and $\theta'$) can be described by exactly the same basic two premises that govern the DMD-quantities for CL and neutrino mass ratios (11) and (15). The choice of the source (in contrast to neutrino mass ratios, it is $\alpha$ not $\alpha_W$) is found from comparison with experimental data – if the dynamical source of lepton mixing is $\alpha_W$ not $\alpha$, the deviations of neutrino mixing angles from maximal value $\pi/4$ would be significantly larger in disagreement with data; similar conclusion follows for the quarks.

So, the two equations for neutrino mixing angles are

$$\mathrm{DMDH}(\theta) \equiv [\mathrm{DMD}(\theta)1]/[\mathrm{DMD}(\theta)2] \cong \alpha(q^2 = M_Z^2),$$

$$[\mathrm{DMD}(\theta)2]^2 = 2[\mathrm{DMD}(\theta)1]. \qquad (23)$$

The explicit form of DMD($\theta$)-quantities can be determined by condition that for small nonzero $\alpha$ the mixing is small deviated from the maximal benchmark one. As a result, the neutrino DMD($\theta$)-quantities (subscript L) are given by

$$[\mathrm{DMD}(\theta)1]_L \equiv \mathrm{Cos}^2(2\theta_{23}), \quad [\mathrm{DMD}(\theta)2]_L \equiv \mathrm{Cos}^2(2\theta_{12}), \qquad (24)$$

they are <1 what justifies the order choice in (23).

The solutions of equations (23)-(24) for neutrino mixing angles are given by

$$\mathrm{Cos}^2(2\theta_{23}) \cong 2\alpha^2(q^2 = M_Z^2), \quad \mathrm{Cos}^2(2\theta_{12}) \cong 2\alpha(q^2 = M_Z^2). \qquad (25)$$

At approximation $\alpha = 0$ neutrinos are 'maximally mixed'; at finite and small actual value of that constant neutrinos are quasi-degenerate with large solar and atmospheric mixing angles and *small but strongly hierarchical deviations from maximal mixing*. Solutions (25) for the neutrino mixing angles reasonably agree with the analysis of available data [8] (quantitative estimations for the neutrino mixing matrix agree with the ones in [5]).

---

[9] As a comment, there is an interesting general analogy between the concept of benchmark flavor pattern and the basic concept of inertial particle motion as a pre-dynamical benchmark concept of motion in Newton classical mechanics where only the deviations from that benchmark motion have dynamical causes.

The quark mixing DMD-quantities follow from the same reasoning that led to neutrino relations (24), but now the realistic mixing should be considered as small deviated from the different type of quark (q) benchmark mixing matrix in (22). And so, instead of neutrino case (24), realistic quark mixing DMD-quantities are

$$[DMD(\theta)2]_q \equiv \sin^2(2\theta_c), \quad [DMD(\theta)1]_q \equiv \sin^2(2\theta'). \tag{26}$$

Notations here are: $\theta_c$ is the Cabibbo mixing angle and $\theta'$ is the next to it smaller one.

Solutions of equations (23) with definition (26) for quark mixing angles are

$$\sin^2(2\theta') \cong 2\alpha^2(q^2 = M_Z^2), \quad \sin^2(2\theta_c) \cong 2\alpha(q^2 = M_Z^2). \tag{27}$$

From comparison of the solutions for neutrino and quark mixing angles, (25) and (27), follows the inference of quark-neutrino mixing angle complementarity [9]:

$$2\theta_{12} \cong \pi/2 - 2\theta_c, \quad 2\theta_{23} \cong \pi/2 - 2\theta'. \tag{28}$$

In the considered model the quark-lepton complementarity condition (28) has a simple physical meaning – it follows from the definition of the primary not dynamical benchmark flavor pattern (22) and the small value of the fine structure constant $\alpha$ being the source (at tree SM approximation) of dynamical shifting of neutrino and quark realistic mixing angles from the extreme benchmark ones.

Note, probably the fact that particle mixing is determined by the dynamical constant $\alpha$, not $\alpha_W$, means that the small universal dynamical particle mixing should be primarily related to Dirac particles - quarks and CL, while the observed large neutrino mixing is secondary[10].

The pairs of neutrino equations (15) and (23)-(25) with the EW charges as sources answer the four topical neutrino questions of a) why the neutrino oscillation solar-atmospheric 3-flavor hierarchy is large, b) why the neutrino mixing is large, c) why neutrino masses should be quasi-degenerate with d) small absolute mass values. The phenomenological answer is: '*since the EW constants $\alpha$ and $\alpha_W$ are small (in comparison with unity), they determine the magnitudes of lepton mass and mixing hierarchies* and the neutrino oscillation mass squared differences are known from experimental data values'.

Dual functions of the small EW charges appear in the present model. These charges are known sources of EW fields, on the one hand, and sources of the particle-copy mass distributions, on the other hand. As a result, the aggregate of free dimensionless flavor and EW parameters in the highly successful three-flavor Electroweak Theory may become a united system.

## 4. Conclusions

Connections between the small EW charges and lepton flavor DMD-quantities are described by three analogous pairs of equations (11), (15) and (23). The common essence of these equations is that they focus two prompted by data physical premises --
i) Causal connections between particle flavor DMDH-hierarchies and EW charges $\alpha$ and $\alpha_W$,
ii) Quadratic DMD-hierarchy relations --
and impart as solutions quantitative observable inferences from these premises.

The main physical inference is that there can be only *two types* of lepton particle-copy groups:

---

[10] It is in conformity with the condition of Dirac-Majorana DMD-duality; it is also a point in favor of Majorana neutrinos.

1) With large DMD-quantities and *large DMD-hierarchy* ~ {$(1/\alpha)$, $(1/\alpha)^2$}; this group contains a pair of large and highly hierarchical mass ratios,

2) With small DMD-quantities and *large DMD-hierarchy* ~ {$\alpha^2_W$, $\alpha_W$}; this group contains quasi-degenerate particles.

From known experimental data definite conclusions follow that the charged leptons belong to the first type, whereas the neutrinos belong to the second type and therefore are quasi-degenerate.

The pair of equations (11) expresses CL mass ratios through the fine structure constant. The pair of equations (15) determines large neutrino oscillation hierarchy and predicts order one QD-neutrino mass ratios; with oscillation data it predicts absolute values of QD-neutrino masses. The pair of equation (23) determines large neutrino mixing as small, but *strongly hierarchical* ~ {$\alpha^2$, $\alpha$}, deviations from maximal benchmark mixing in agreement with [5].

As it appears from the above discussion, the quark-lepton mixing complementarity [9] has a clear physical meaning – it is the result of adding the one small universal dynamical mixing pattern (probably of Dirac particles - quarks and CL) to the necessary two originally extremely different not dynamical quark and neutrino (probably Majorana) benchmark mixing patterns.

## Appendix. Benchmark flavor pattern

The concept of benchmark flavor pattern, as a primary not dynamical (independent of the interaction constants of the Standard Model) pattern of elementary particle dimensionless flavor quantities, is suggested in ref.[5] and shown that it is relevant for phenomenological explanation of the realistic flavor patterns of charged leptons, neutrinos and quarks. In this Appendix we consider that concept as the zero approximation in the electroweak interaction constants $\alpha$ and $\alpha_W$ serving as a benchmark pattern for the realistic lepton flavor mass and mixing quantities.

It makes sense to include in the lepton benchmark flavor pattern the quadratic DMD-hierarchy equations (7) to have all relations, which are independent of dynamical parameters, singled out to the benchmark system. So by definition the lepton benchmark flavor pattern is given by:

$$\alpha(\alpha_W) = 0 - \text{no interactions}, \quad (A1)$$

$$m_e \cong (m_e)_{exp}, \quad m_\nu \cong 0, \quad m_\mu, \quad m_\tau \cong \infty, \quad (A2)$$

$$DMD(\nu)1 = DMD(\nu)2 = 0, \quad DMD(CL)1 = DMD(CL)2 = \infty, \quad (A3)$$

$$[DMD(1)]^2 / [DMD(2)]_{CL} = 2, \quad [DMD(2)]^2 / [DMD(1)]_\nu = 2, \quad (A4)$$

$$\begin{pmatrix} 1/\sqrt{2} & 1/\sqrt{2} & 0 \\ -1/2 & 1/2 & 1/\sqrt{2} \\ 1/2 & -1/2 & 1/\sqrt{2} \end{pmatrix} \nu. \quad (A5)$$

The neutrino benchmark mixing matrix (A5) has the known bimaximal form[11].

From the quadratic lepton benchmark hierarchies (A4) it follows that the linear lepton DMDH-hierarchies defined in Eq. (2) and (4) are equal zero at benchmark pattern:

$$[DMD(1)] / [DMD(2)]_{CL} = 0, \quad [DMD(1)] / [DMD(2)]_\nu = 0. \quad (A6)$$

It is since, from (A3), the benchmark CL DMD-quantities are infinitely large and the neutrino ones are equal zero.

The two lepton flavor equations (6) and (7) in the main text appear from the benchmark ones (A4) and (A6) after emergence of the small finite parameters $\alpha$ and $\alpha_W$ instead of zeros in Eq.(A6). Then, finite realistic lepton flavor quantities (12), (13) and (16), (17) appear from the benchmark ones (A3) as solutions of the equations (11) and (15) in the main text. Note that from (A2) the relation $m_\nu / m_e \cong 0$ follows at benchmark pattern; so in accord with the reasoning in [5], after the emergence of small $\alpha$-parameter, QD-neutrinos should appear with mass scale given by

$$m_\nu \cong \pi \alpha^3 m_e / 3 \cong 0.2 \text{ eV}. \quad (A7)$$

The benchmark lepton flavor pattern described by equations (A1)-(A6) is an idealized concept with no explicit relation to EW interactions. The realistic lepton physics appears as deviated from the benchmark one by small shifting from zero values of the EW constants $\alpha$ and $\alpha_W$ in the linear DMDH-hierarchy relations (A6); that small deviation from benchmark pattern determines very large for CL and very small for neutrinos shifting of lepton benchmark masses and DMD-quantities (A2) and (A3) to the realistic values.

---

[11] Bimaximal neutrino mixing was widely discussed in the literature (see e.g. [10]) as a symmetrical approximate description of the large neutrino mixing. Here it is considered as pre-dynamical neutrino (probably Majorana) benchmark maximal mixing level from which the deviation, caused by emergence of small dynamical $\alpha$-parameter, is measured.

To conclude, the main idea of this paper, as described by Eq.(5), is indeed an inference from the quadratic DMD-hierarchy equations (A4) of the benchmark pattern (Eq.(7) in the main text) and the emergence of small dynamical EW constants $\alpha, \alpha_W \ll 1$ in realistic lepton physics with three flavor generations. Thus, realistic finite lepton masses and large mass and mixing DMD-hierarchies appear from the benchmark pattern (A1)–(A6) by emergence of *small EW charges as sources of the necessary deviation from that pattern.*